\documentclass[aps,prl,twocolumn,showpacs,preprintnumbers,superscriptaddress]{revtex4}

\usepackage[pdftex]{graphicx}

\input def.sty

\usepackage{everypage}
\usepackage{gensymb}

\begin{document}
\title{New Precision Limit on the Strange Vector Form Factors of the Proton}
\date{\today}%

\begin{abstract}
The parity-violating cross-section asymmetry in the elastic scattering of polarized electrons from unpolarized protons 
has been measured at a four-momentum transfer squared $\qsq = 0.624\gevc$ and beam energy 
$E_{b}=3.48~\mbox{GeV}$ to be  $\APV = -23.80 \pm 0.78 \text{(stat)} \pm 0.36 \text{(syst)}$~parts per million.  
This result is consistent with zero contribution of strange quarks to the combination of electric and magnetic form factors $G_E^s + 0.517~G_M^s = 0.003 \pm 0.010\text{(stat)} \pm 0.004\text{(syst)} \pm 0.009\text{(f\,f)}$, where the third error is due to the limits of precision on the electromagnetic form factors and radiative corrections. With this measurement, the world data on strange contributions to nucleon form factors are seen to be consistent with zero and not more than a few percent of the proton form factors.
\end{abstract}

%
%
%
\collaboration{The HAPPEX Collaboration}
\noaffiliation
\author{Z.~Ahmed}
\affiliation{Syracuse University, Syracuse, New York 13244, USA} 

\author{ K.~Allada}
\affiliation{University of Kentucky, Lexington, Kentucky 40506, USA}

\author{K.~A.~Aniol}
\affiliation{ \mbox{California State University, Los Angeles}, Los Angeles, California 90032, USA }

\author{D.~S.~Armstrong}
\affiliation{College of William and Mary, Williamsburg, Virginia 23187, USA}

\author{J.~Arrington}
\affiliation{Argonne National Laboratory, Argonne, Illinois, 60439, USA}

\author{ P.~Baturin}
\affiliation{Florida International University, Miami, Florida 33199, USA}

\author{ V.~Bellini}
\affiliation{Istituto Nazionale di Fisica Nucleare, Dipt.~di Fisica dell'Univ.~di Catania, I-95123 Catania, Italy}

\author{J.~Benesch}
\affiliation{Thomas Jefferson National Accelerator Facility, Newport News, Virginia 23606, USA} 

\author{R.~Beminiwattha}
\affiliation{Ohio University, Athens, Ohio 45701, USA}

\author{F.~Benmokhtar}
\affiliation{Carnegie Mellon University, Pittsburgh, Pennsylvania 15213, USA}

% ...


\maketitle

It has long been established that a complete characterization of nucleon substructure must go beyond three valence quarks and include the $q\bar{q}$ sea and gluons.
In deep inelastic scattering, for example, sea quarks are known to dominate interactions in 
certain kinematic regimes. With the discovery by the EMC collaboration~\cite{Ashman:1987hv} that quark spins are not the 
dominant contribution to nucleon spin, the role of sea quarks, and especially strange quarks, has been scrutinized.  
More generally, since valence-quark masses account for only about 1\% of the nucleon mass, 
a better understanding of the role of gluons and sea quarks in nucleon substructure is imperative.
Cleanly isolating the effects of the 
quark sea is typically difficult; one notable exception is the extraction of 
the vector strange matrix elements $\langle \overline s \gamma_\mu s\rangle$ in semi-leptonic neutral weak 
scattering~\cite{Kaplan:1988ku}.

A quantitative understanding of the role of strange quarks in the nucleon would have broad implications. 
The range of uncertainty in the strange-quark condensate $\langle \overline s s\rangle$ leads to an order
of magnitude uncertainty in spin-independent scattering rates of dark matter candidates,
while spin-dependent rates are uncertain to a factor of two given the range of uncertainty in the strange-quark contribution to nucleon spin, $\Delta s$~\cite{Ellis:2008hf}.  
The strange-sea asymmetry $s-\overline s$ is important for the interpretation of the NuTeV experiment~\cite{Lai:2007dq,Olness:2003wz}.  
A better understanding of strangeness in the nucleon will clarify issues for many specific experiments as well as improve
our understanding of the role of sea quarks in general.

Following the recognition that parity-violating electron scattering can measure the neutral weak form factors and 
hence the vector strange-quark matrix elements~\cite{McKeown:1989ir}, numerous experiments have been performed.  
Several such experiments presented evidence supporting non-zero strange form factors, although the significance of the effect was limited~\cite{Armstrong:2005hs,Maas:2004ta,Maas:2004dh}.  
In contrast, the HAPPEX collaboration has found results consistent with zero strangeness in each of several measurements at various values of the four-momentum transfer squared $\qsq$ ~\cite{Aniol:2004hp,Acha:2006my}.  
The HAPPEX measurements, while only capable of measuring a single value of $\qsq$ at a time,  have put particular emphasis on high statistical accuracy and small systematic uncertainties.

In this paper, we report a new measurement performed in Hall A at Jefferson Laboratory.  The 
kinematics of the measurement were chosen to be particularly sensitive to the apparent effects reported in~\cite{Armstrong:2005hs}. 
The experimental technique was similar to previous HAPPEX  measurements~\cite{Aniol:2004hp}. A 100~$\mu$A continuous electron beam of longitudinally polarized electrons at 3.481~GeV was incident on a 25~cm long liquid hydrogen target.  
The twin Hall A High Resolution Spectrometers (HRS)~\cite{Alcorn:2004sb} each accepted scattered electrons over a solid angle of 5~msr with an averaged polar angle of $\langle \theta \rangle \sim 13.7 \degree$.  Electrons which scattered elastically from protons were focused onto a calorimeter in each spectrometer; electrons from inelastic processes on free protons were not transported to the focal plane. 
Each calorimeter was composed of alternating layers of lead and lucite, with \v{C}erenkov light from the electromagnetic shower collected by a single photomultiplier tube.  

The polarized beam is generated through photoemission from a doped GaAs superlattice crystal. 
The polarization state of the electron beam was held constant for a time window of about 33~ms, then flipped to the complementary state. The polarities of these pairs of time windows were selected from a pseudorandom sequence. 
The responses of beam monitors and the electron calorimeters were integrated over each period of stable helicity. 
Periods of instability in the beam, spectrometer, or data acquisition electronics were cut from the accepted data.
A total of $29.9 \times 10^6$ pairs passed all cuts and formed the final data sample, including $1.0 \times 
10^6$ pairs in which only one of the two spectrometers was functional.

The helicity-dependent asymmetry in the integrated calorimeter response $\ARAW$ was computed for each pair of helicity windows.
The physics asymmetry $\APV$ is derived after normalization for beam intensity fluctuations, with corrections for background contributions, kinematics normalization, beam polarization, and changes in beam energy and trajectory.   
The magnitude and estimated uncertainty due to each of these corrections is described below and summarized in Table~\ref{tab:errorsummary}.  

The laser optics of the polarized source were carefully configured to minimize changes to the electron beam parameters under polarization 
reversal~\cite{Paschke:2007zz}. 
A feedback system 
was used to minimize the helicity-correlated intensity asymmetry of the beam. Averaged over the course of the experimental run, the helicity-correlated asymmetries in the electron beam were $0.20$ parts per million (ppm) in intensity, $0.003$~ppm in energy, and $3$~nm in position.

Due to the symmetric acceptance of the two spectrometers and the small run-averaged values of 
helicity-correlated beam asymmetries, the cumulative correction due to beam trajectory and energy asymmetry was only 
$0.016\pm0.034$~ppm.  
The calorimeter system response was measured to be linear, with an uncertainty of less than $0.5\%$, through dedicated tests using pulsed LEDs.

Electrons scattered from the aluminum windows of the cryogenic hydrogen vessel were the largest background.  
Due to the high $\qsq$, aluminum elastic scattering did not contribute significantly, leaving 
quasielastic scattering as the dominant background source. The contributed signal fraction 
was determined to be $(1.15 \pm 0.35)\%$ using the evacuated target cell to directly measure the aluminum-scattered rate; 
these rates were checked using aluminum targets matched to the full target radiation length.  
The asymmetry of this background was calculated to be $-34.5$~ppm, 
with an uncertainty of 30\% to account for potential contributions from inelastic states.  

Inelastically scattered electrons can also rescatter in the spectrometer and produce a signal in the calorimeter. 
Dedicated studies of 
electron rescattering in the spectrometer were combined with parameterizations of the electron-proton inelastic spectra to estimate a fractional 
contribution of $(0.29 \pm 0.08)\%$ to the total rate.  The dominant mechanism was $\Delta$ production, for which the theoretical 
calculated asymmetry of $-63$~ppm was used with an uncertainty of 20\%. An additional systematic uncertainty contribution of 0.14~ppm  accounted for the possibility that a small fraction of the signal ($<10^{-4}$) could have originated from rescattering with ferromagnetic material~\cite{Aniol:2004hp}.
The total correction from all sources of background amounted to $(1.0 \pm 0.8)\%$ of $\APV$.

Both Compton and M{\o}ller scattering processes were used to precisely determine the electron beam polarization.  
The accuracy of the Hall A M{\o}ller polarimeter was improved through a careful study of the 
uniformity of the ferromagnetic foil target, leading to a result of  $(89.2\pm1.5)\%$.
The dominant source of uncertainty in previous analyses of backscattered photons in the Hall A Compton polarimeter~\cite{Alcorn:2004sb} lay in the effect of the trigger threshold on the normalization of the analyzing power.   
This was improved through threshold-less integration of the photon signal, with a result of $(89.41\pm0.86)\%$.
Averaged, the beam polarization was determined to be $(89.36 \pm 0.75)\%$.

Dedicated low-current data were periodically taken to measure $Q^2$ using the standard tracking package of the HRS~\cite{Alcorn:2004sb}. 
A water target was used to calibrate the spectrometer angle, with momentum differences from the elastic hydrogen and elastic and inelastic oxygen peaks determining the scattering angle to a precision of 0.4~mrad.  
Including the spectrometer calibration resolution, the average $Q^2$ was determined to be $0.624 \pm 0.003 \gevc$, which implies a 0.8\% uncertainty on the quoted $\APV$. 
An additional correction factor $\kappa$, which relates the asymmetry measurement over a finite range of initial-state energy and solid angle to the quoted $\qsq$,  was determined through simulation to be $\kappa = 0.995\pm0.002$. 

After all corrections to $\ARAW$, as summarized in Table~\ref{tab:errorsummary}, the parity-violating asymmetry $\APV = -23.80 \pm 0.78~\text{(stat)} \pm 0.36~\text{(syst)}$~ppm at $\qsq=0.624\gevc$. 

\begin{table}
\centering
\begin{tabular}{lr}
\hline
\multicolumn{2}{c}{$\ARAW = -21.78 \pm 0.69$~\mbox{ppm}} \\ \hline \hline
Detector Linearity & $0.0\% \pm  0.5\%$ \\ \hline
Beam Asymmetries  &  $-0.9\%   \pm 0.2\%$ \\ \hline
Backgrounds & $-1.0\%  \pm 0.8\%$ \\ \hline 
Acceptance Factor $\kappa$ &   $-0.5\%  \pm 0.2\%$\\ \hline
Beam Polarization   &  $10.9\%  \pm 0.8\%$ \\ \hline 
 $Q^2$             &    $\mbox{\hspace{0.2cm} -- \hspace{0.3cm}}  \pm 0.8\%$ \\ \hline \hline
{\bf Total} & $8.5\% \pm 1.5\%$ \\ \hline \hline
\multicolumn{2}{c}{$\APV = -23.80 \pm 0.78 \pm 0.36$~\mbox{ppm}} \\ \hline
\end{tabular}
\caption{Summary of corrections to the raw asymmetry and the associated systematic uncertainty estimates as a fraction of $\APV$. The uncertainty on $\ARAW$ is statistical only, while $\APV$ is listed with statistical and experimental systematic errors. }
\label{tab:errorsummary}
\end{table}

Following notation from~\cite{Maas:2004dh}, the theoretical expectation for $\APV$ can be expressed in three terms: $\APV = A_V + A_A + A_S$.  
$A_V$ and $A_A$ depend on the proton weak charge $(1-4\sin^2\theta_W)$ and the nucleon vector 
and axial-vector electromagnetic form factors, respectively, while strange-quark contributions to 
the vector form factors are isolated in $A_S$.  At tree level,
\begin{equation}
A_S = A_0  \left[ \frac{\epsilon G^p_E G^s_E  + \tau G^p_M G^s_M}{\epsilon (G^p_E)^2 + \tau (G^p_M)^2}   \right].
\end{equation}
Here $A_0 = G_F \qsq / ( 2 \pi \sqrt{2} \alpha )$, $\tau = \qsq/(4M_p)$, $\epsilon = \left[ 1 + 2(1+\tau)\tan^2 ( \theta / 2 )\right]^{-1}$, and $G^p_{E(M)}$ is the proton electric (magnetic) form factor.

 \begin{figure}\begin{center}
 \includegraphics[width=3.6in,angle=0]{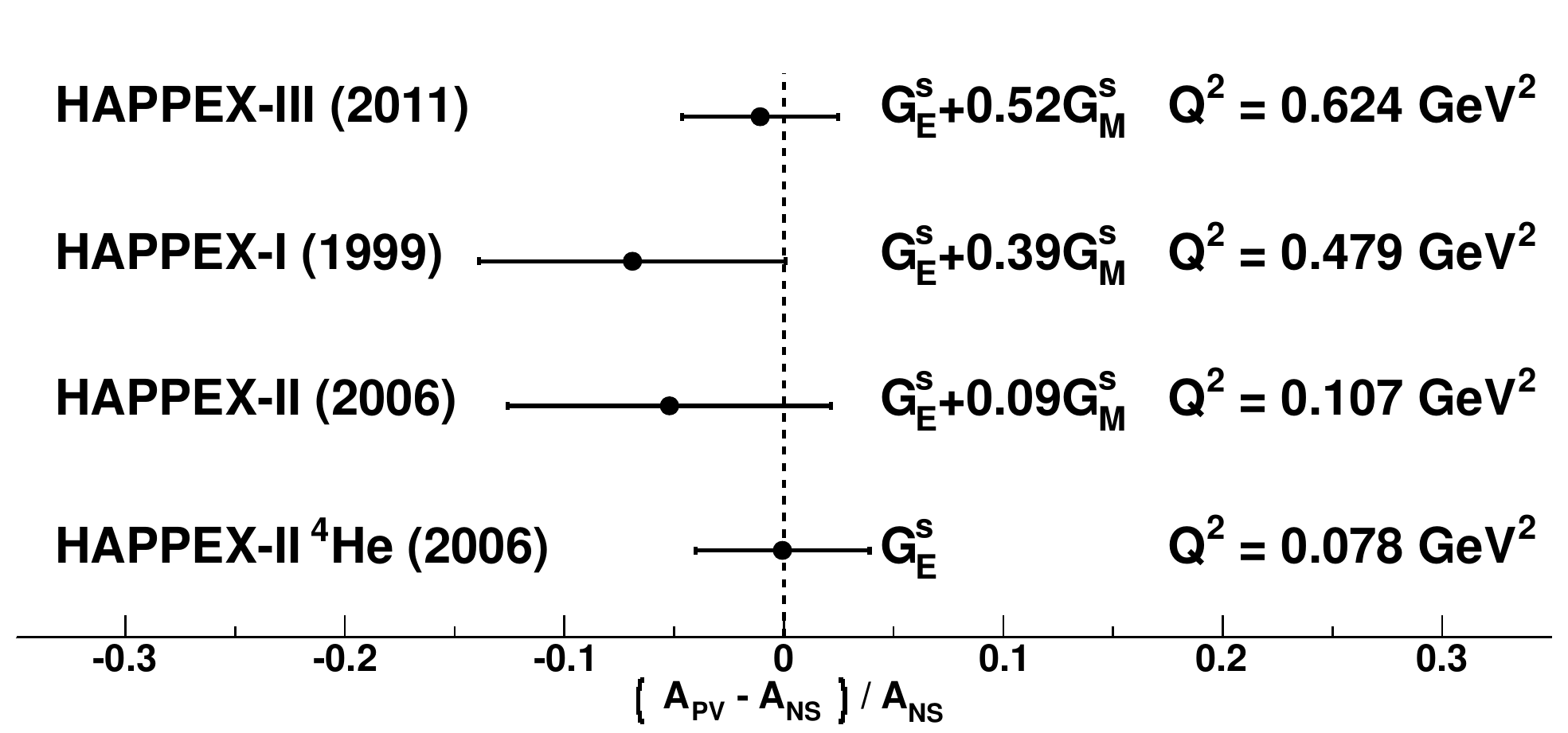}
 \caption{$(\APV - \ANS ) / \ANS$ for the four HAPPEX measurements (this result and~\cite{Aniol:2004hp,Acha:2006my}). The linear combination of form factors to which each measurement is sensitive is listed.  For each measurement, results are consistent with zero strangeness contribution.}
 \label{fig:hapcompare}
 \end{center}\end{figure}

If strange quarks did not contribute to the vector form factors, the asymmetry at $\langle \qsq \rangle = 0.624\gevc$ would be expected to be $\ANS = A_V + A_A = -24.062 \pm 0.734$~ppm. 
This calculation utilizes parameterizations of the electromagnetic form factors which incorporate 
two-photon-exchange corrections to published form-factor data~\cite{Arrington:2006hm}. 
The uncertainty in $\ANS$ primarily results from uncertainties in these form factors and in radiative corrections in the axial term $A_A$  involving parity-violating multi-quark interactions.  
While theoretical investigation~\cite{Zhu:2000gn}  has suggested that the latter corrections could be as large as 30\% of the axial form factor, the net uncertainty in $\ANS$ is small for forward-angle studies where the small coefficient $\sqrt{1-\epsilon^2} (1-4\sin^2\theta_W) $ suppresses the axial term.  
The uncertainty in these corrections, as a fraction of the axial form factor, is assumed to be constant with $\qsq$.
Standard electroweak corrections~\cite{Nakamura:2010zzi} are also included in $\ANS$ and contribute negligible uncertainty.  
Additional radiative corrections involving two-photon exchange, expected to be at the level of 0.03\%~\cite{Arrington:2006hm}, are neglected.
Comparing $\ANS$ to the measured $\APV$, the strange-quark contributions are determined to be 
$\ges + 0.517~\gms = 0.003 \pm 0.010 \pm 0.004 \pm 0.009$, where the error bars correspond to statistical, systematic, and the $\ANS$ uncertainties, respectively.

Our result is compared to previous measurements from the HAPPEX collaboration in Fig.~\ref{fig:hapcompare}, which displays  $(\APV - \ANS)/\ANS$, 
the fractional deviation from theoretical expectation in the absence of strange-quark contributions.  
The $\qsq$ and specific form-factor sensitivity for each measurement are noted on the figure.  The error bars include experimental uncertainties, while the uncertainty in $\ANS$ is taken to be zero for this plot.
For forward angle scattering at each indicated $\qsq$, the HAPPEX measurements represent the most accurate determinations of the strange vector matrix elements; they show no indication of a signal for strange-quark contributions to the form factors. 

\begin{figure}\begin{center}
    \includegraphics[height=2.6in,angle=0]{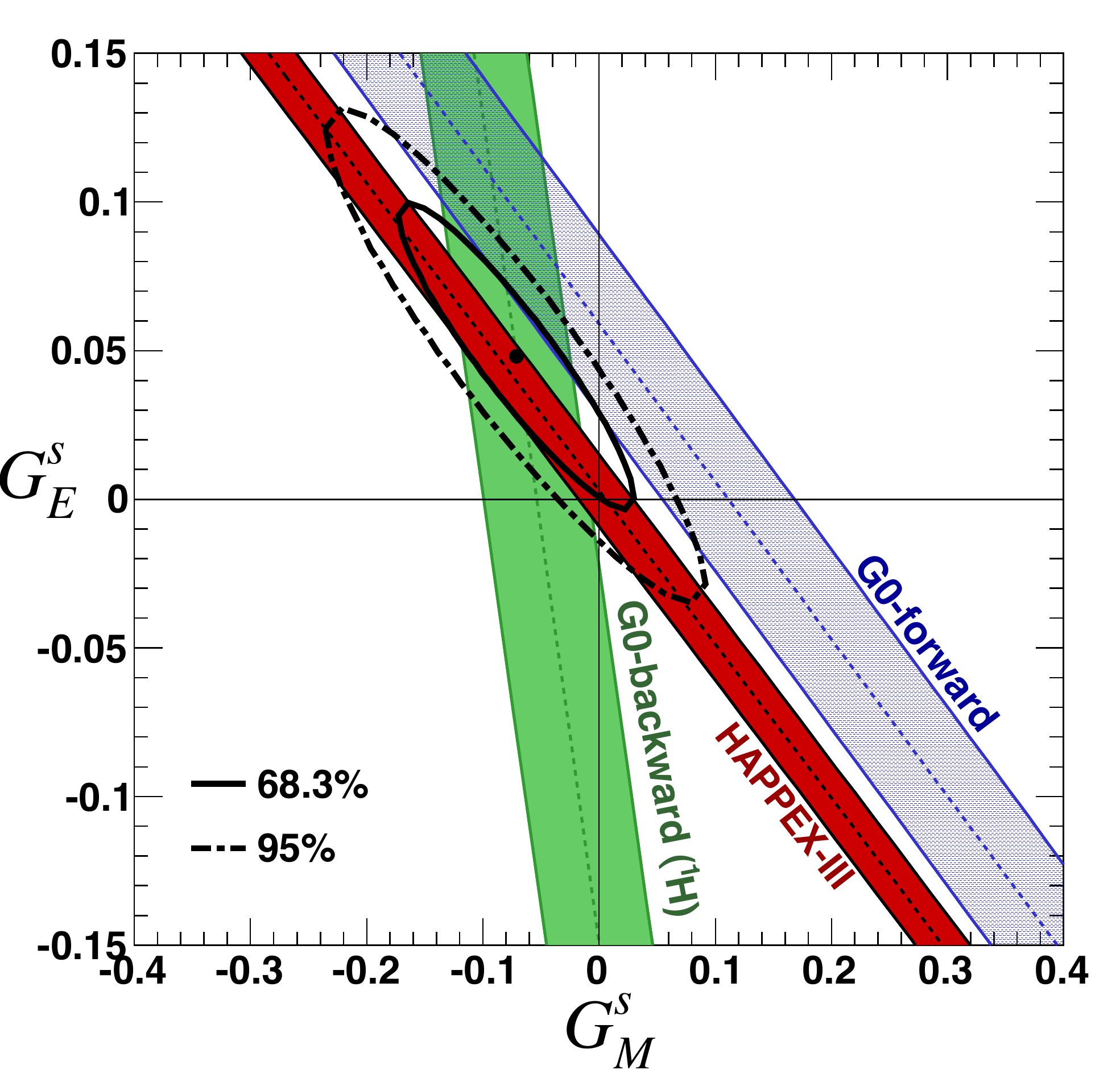}
\caption{Constraints on $G_E^s$ and $G_M^s$ at $\qsq\sim0.62\gevc$.  The experimental bands are from the results presented in this letter (HAPPEX-III) and the G0 
measurements~\protect{\cite{Armstrong:2005hs, Androic:2009zu}}.} 
\label{fig:contour}
\end{center}\end{figure}

The constraints on the 2-D space spanned by $\ges$ and $\gms$ from all measurements near $\qsq\sim0.62\gevc$ are shown in Fig.~\ref{fig:contour}.
The experimental constraints at 1$\sigma$ are represented by the shaded bands 
indicating the combined statistical and experimental systematic error bars.  
The contours, representing the 68\% and 95\% uncertainty boundaries as indicated, combine all three measurements and also account for the uncertainties in $\ANS$.
The independently separated values resulting from this fit are  $G_E^s= 0.047 \pm 0.034$ and 
$G_M^s = -0.070 \pm 0.067$, with a correlation coefficient of $-0.93$. 
The combined constraint is consistent with $\ges = \gms = 0$.

\begin{figure}\begin{center}
\hspace*{-0.3in}\includegraphics[width=4.0in,angle=0]{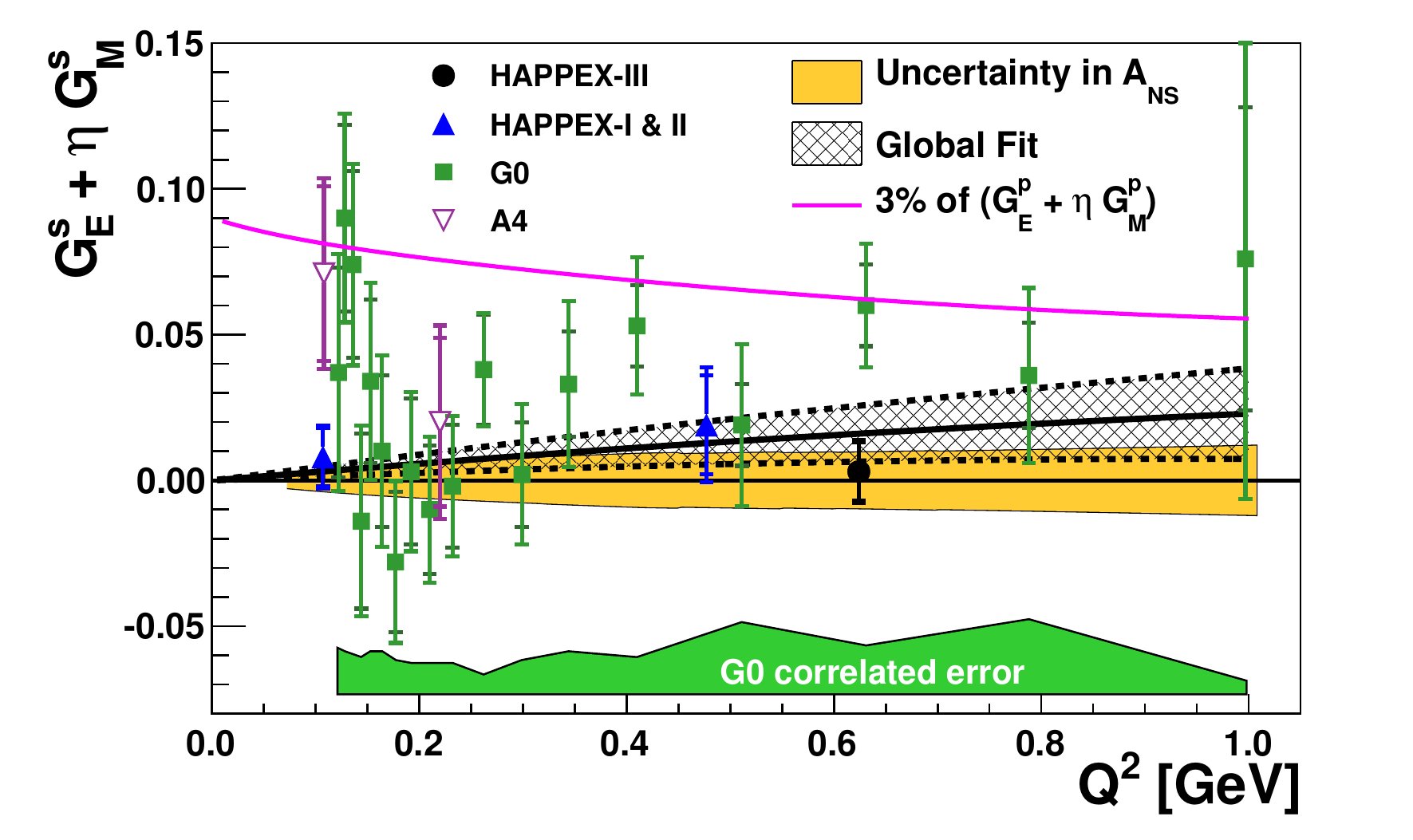}
\caption{ Results of strange-quark vector form factors for all measurements of forward-angle scattering from the proton. The global fit is described in the text. The solid curve represents a 3\% contribution to the comparable linear combination of proton form factors.  
}
\label{fig:GeGm}
\end{center}\end{figure}

Figure~\ref{fig:GeGm} shows all published data on the net strangeness contribution $G_E^s + \eta G_M^s$ in forward-angle scattering 
measurements from the proton  versus $\qsq$.  Here, $\eta = \tau G_M^p / (\epsilon G_E^p)$, and is approximately numerically equal to 
$\qsq$/($\gevc$) over the range of the plot.  Data from the HAPPEX~\cite{Aniol:2004hp,Acha:2006my}, G0~\cite{Armstrong:2005hs}, and 
A4~\cite{Maas:2004ta,Maas:2004dh} collaborations are shown.  On each data point, the error bars indicate both the statistical error and the 
quadrature sum of statistical and uncorrelated systematic error.  
For the G0 data, some systematic uncertainties are correlated between points with a magnitude indicated by the shaded region at the bottom of the plot.
A shaded region around the zero-net-strangeness line represents the uncertainties in $\ANS$ at $1\sigma$; this uncertainty is not also included in the individual data points. 

While there is no reliable theoretical guidance on the possible $\qsq$-dependence of the strange form factors, it is reasonable to expect that they would not change rapidly with $\qsq$, consistent with nucleon form factors in this range which are described to a reasonable precision by smooth dipole or Galster parameterizations~\cite{Arrington:2006hm}. 
The cross-hatched region displays the 1$\sigma$ region allowed by a leading-order fit in which $\gms$ is taken to be constant and $\ges$ is proportional to $\qsq$.    This parameterization follows that of~\cite{Young:2006jc,Liu:2007yi}.
The fit includes all published data, 
including HAPPEX-II $^{4}$He~\cite{Acha:2006my} and backward-angle proton 
measurements~\cite{Androic:2009zu,Spayde:2003nr,Baunack:2009gy}, and takes the correlated uncertainties in the G0 forward-angle data into account but neglects the uncertainty in $\ANS$.  
The confidence level of the fit is 33\%, demonstrating the reasonable self-consistency of the data.  
In contrast to the situation prior to this work, the HAPPEX-III point constrains the cross-hatched fit into significant overlap with the band corresponding to the uncertainty in $\ANS$, and just over $1\sigma$ from zero.
Fits using alternative parameterizations ({\it e.g.\ } next higher power in $\qsq$, dipole and Galster forms for $\gms$ and $\ges$) similarly remain consistent with zero. 
Thus, the results of this letter rule out large contributions from strange vector form factors with $\qsq$ behavior similar to that of the nucleon electromagnetic form factors.

\begin{acknowledgments}

This work was supported by the U.S. Department of Energy and National Science Foundation. 
Jefferson Science Associates, LLC,  operates Jefferson Lab for the U.S. DOE under U.S. DOE 
contract DE-AC05-060R23177.

\end{acknowledgments}

\end{document}